\documentclass[pra,twocolumn]{revtex4-1}
\usepackage{bm, amssymb, physics, environ, tikz, float, amsmath, subcaption}
\usetikzlibrary{decorations.pathmorphing, shapes.misc, shapes.geometric,calc, arrows.meta, arrows, decorations.pathmorphing,fit}
\definecolor{Dark_Green}{rgb}{0,128,0}

\begin{document}
\title{Electromagnetically induced transparency in many-emitter waveguide quantum electrodynamics: linear versus nonlinear waveguide dispersions}
\author{Tiberius Berndsen and Imran M. Mirza} 
\email{mirzaim@miamioh.edu}
\affiliation{Macklin Quantum Information Sciences, Department of Physics, Miami University, Oxford, Ohio 45056, USA}

\begin{abstract}
We study single-photon induced electromagnetically induced transparency (EIT) in many-emitter waveguide quantum electrodynamics (wQED) with linear and nonlinear waveguide dispersion relations. In the single-emitter problem, in addition to the robustness of the EIT spectral features in the over-coupled regime of wQED, we find that the nonlinear dispersion results in the appearance of a side peak for frequencies smaller than the resonant EIT frequency which turns into a pronounced plateau as the nonlinearity is enhanced. Consequently, for many-emitter scenarios, our results indicate the formation of band structure which for higher values of nonlinearity leads to narrow band gaps as compared to the corresponding linear dispersion case. Long-distance quantum networking aided with quantum memories can serve as one of the targeted applications of this work.  
\end{abstract}

\maketitle

{\it Introduction.}---EIT is a coherent optical phenomenon where, under the right conditions of resonant transitions and atomic decay rates, a weak probe light beam, in the presence of a strong pump field, can pass through a three-level atomic medium with 100\% transmission rate \cite{boller1991observation,fleischhauer2005electromagnetically}. In addition to providing this remarkable transparency feature, EIT also allows the manipulation of dispersion properties of the probe field leading to fascinating effects such as slow, stopped, and fast light \cite{kocharovskaya2001stopping,safavi2011electromagnetically, walsworth2002story}. These two features combined, EIT in the last two decades has witnessed a range of applications from quantum information storage \cite{lvovsky2009optical, beausoleil2004applications} to magnetometry \cite{kitching2018chip} to the development of more precise atomic clocks \cite{vanier2005atomic}.

In this work, we study EIT in wQED - quantum emitters/QEs (natural or artificial atoms, quantum dots, or qubits) coupled to traveling optical fields guided by 1D bosonic fibers or waveguides \cite{sheremet2023waveguide, roy2017colloquium, xiang2013hybrid}. In its simplest setting, a wQED setup consists of a single two-level QE coupled with a bidirectional waveguide with linear dispersive properties \cite{shen2005coherent}. However, for the observation of EIT, one requires to utilize a three-level QE in which two atomic transitions, initiated from two different energy levels, are guided to a common final state \cite{mirza2018influence}.

To date, the literature on the topic of EIT in wQED has remained limited to either the case in which a chain of three-level atoms coupled to a waveguide with linear dispersion \cite{mirza2018influence, fang2016photon, bermel2006single} or a single three-level atom coupled to a nonlinear waveguide \cite{martens2013photon, gong2008controlling}. Relevant to this are also the studies in which single or many but two-level QEs interact with a nonlinear dispersive waveguide \cite{zhou2008controllable, schneider2016green, werra2013spectra, zang2010single, roy2011correlated} or strong interaction at the few-photon level has been established in a waveguide with a nonlinear medium \cite{hafezi2012quantum, firstenberg2013attractive}. However, none of the reported work, to the best of our knowledge, has examined the problem of EIT in the case of a 1D lattice of many three-level QEs coupled to a waveguide with nonlinear dispersion relation. Keeping in view the rich many-body physics involved in the scenario of a one-dimensional periodic lattice of QEs coupled to guided photonic modes \cite{fayard2021many, poshakinskiy2021dimerization, mirza2017chirality} and the novel interplay of the EIT with nonlinear waveguide dispersion, in this letter we address this problem. 

As some of the key results, we find that for single emitter problems, the nonlinear dispersion relation can result in the appearance of an additional side peak at smaller frequencies (compared to the EIT peak frequency) which forms the shape of a transmission plateau for stronger nonlinear dispersion relations. For many-emitter lattices, by utilizing the transfer matrices framework \cite{linton2009wave, mostafazadeh2020transfer}, we observe that due to quantum interference, the photon transmission properties exhibit band structures whose gap tends to reduce with elevated nonlinear dispersion. Our results indicate that the nonlinear dispersion of the waveguide allows novel ways of controlling the single-photon propagation in three-level wQED setups with potential applications in long-distance quantum networking and communications \cite{kimble2008quantum}.

The paper is structured as follows. We begin by recapping the problem of single-photon transport through a three-level wQED setup with linear dispersion \cite{witthaut2010photon, mirza2018influence}. Next, we examine the same problem for nonlinear dispersive waveguides and discuss the novel problem of periodic chains of three-level QEs while comparing the transmission spectra and dispersion curves for the linear and nonlinear cases. Finally, we close with the main conclusions.
\begin{figure*}
\hspace{-2mm}\begin{subfigure}[b]{0.45\linewidth}
\includegraphics[width=2.85in, height=1.05in]{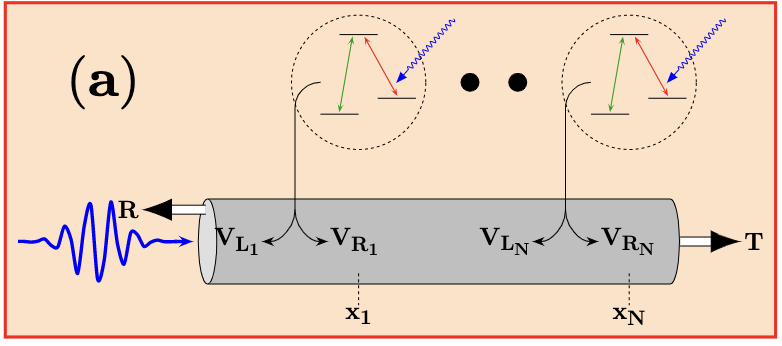}
\\
\includegraphics[width=2.85in, height=0.50in]{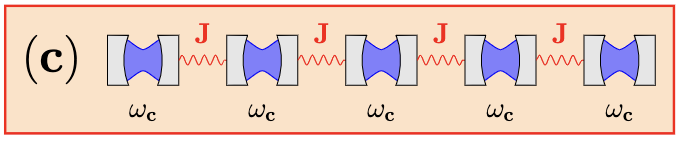}   
  \end{subfigure}  
  \hspace{-4.25mm}
\begin{subfigure}[b]{0.27\linewidth}
\includegraphics[width=1.9in, height=1.4in]{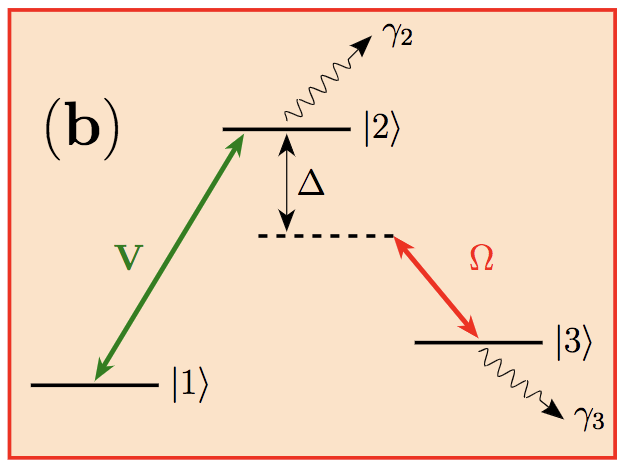}
\end{subfigure}%
\hspace{-0.75mm}
\begin{subfigure}[b]{0.27\linewidth}
\includegraphics[width=1.8in, height=1.1in]{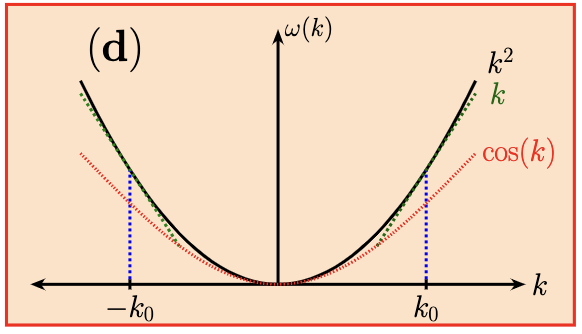}
\end{subfigure}%
\captionsetup{
format=plain,
margin=1em,
justification=raggedright,
singlelinecheck=false
}
\caption{{\bf (a) Setup:} A bidirectional waveguide with linear or nonlinear dispersion is side coupled to a chain of three-level QEs, with left (right) emitter-waveguide coupling strength $V_{L_j}$($V_{R_j}$) for the $j$th QE ($j=1,2,..., N$). Each QE is excited by a pump field and an incoming single photon. {\bf (b) Energy-level diagram of the three-level QE:} Emitter-waveguide coupling drives the transition $\ket{1}\longleftrightarrow\ket{2}$ with coupling strength $V$ (here $V_L=V_R=V$) while the transition $\ket{2}\longleftrightarrow \ket{3}$ is driven by a strong pump field with Rabi frequency $\Omega$ and detuning $\Delta$. $\Delta$ has been exaggerated for visual clarity. The decay rates are $\gamma_{2}$ and $\gamma_{3}$ from the states $\ket{2}$ and $\ket{3}$, respectively. {\bf (c) Nonlinear waveguide model:} Waveguide here consists of a 1D array of coupled resonators with an identical frequency $\omega_c$ and a photon hopping rate $J$ between any two consecutive resonators. Under the tight-binding approximation, such a setup is known to offer a cosine-like nonlinear dispersion for photon propagation. {\bf (d) Approximations of the nonlinear dispersion:} Quadratic and linear approximations of a cosine dispersion relation are shown. For simplicity, the lattice constant has been set equal to unity.} \label{Schematic_Diagram}
\end{figure*}
{\it Linear Dispersion case: Single QE problem.}---As shown in Fig.~\ref{Schematic_Diagram}(a), the Hamiltonian for the system under study can be decomposed into three pieces
\vspace{-1mm}
\begin{align}\label{HamLin}
\hat{\mathcal{H}} = \hat{\mathcal{H}}_{W} + \hat{\mathcal{H}}_{QE} + \hat{\mathcal{H}}_{I},
\end{align}
where $\hat{\mathcal{H}}_{W}$, $\hat{\mathcal{H}}_{QE}$, and $\hat{\mathcal{H}}_{I}$ represent the free waveguide, free three-level QE, and interaction Hamiltonian, respectively. We first focus on our three-level $\Lambda$-type QE Hamiltonian \cite{sen2015comparison}. As mentioned in Fig.~\ref{Schematic_Diagram}(b), the QE Hamiltonian can be expressed as (with $\hbar=1$) 
\begin{align}
\hat{\mathcal{H}}_{QE}= \widetilde{\omega}_{2}\ket{2}\bra{2} + \widetilde{\omega}_{3}\ket{3}\bra{3}+ \frac{\Omega}{2}\left(\ket{3}\bra{2} + h.c.\right), 
\end{align}
where we have adopted a short notation in which $\widetilde{\omega}_2\equiv\omega_{2}-\frac{i\gamma_{2}}{2}$, $\widetilde{\omega}_3\equiv\omega_{2} - \Delta -\frac{i\gamma_{3}}{2}$, and h.c. abbreviates the hermitian conjugate of the first term in the parenthesis. In the real-space formalism of quantum optics \cite{shen2009theoryI, shen2009theoryII}, the waveguide Hamiltonian in the linear dispersion regime takes the following form
\vspace{-2mm}
\begin{align}\label{Hamiltonian_Linear}
\hat{\mathcal{H}}^{(l)}_{W} = iv_{g}\int\big( \hat{c}^{\dagger}_{L}\partial_{x}\hat{c}_{L} - \hat{c}^{\dagger}_{R}\partial_{x}\hat{c}_{R} \big)dx,
\end{align}
with $v_{g}$ being the group velocity of the photon in the waveguide. Finally, under the rotating wave approximation, we write the interaction Hamiltonian in the following fashion
\begin{align}
\hat{\mathcal{H}}_{I}= \sum_{d=L,R}\int\delta(x)V_{d}\left(\hat{c}^{\dagger}_{d}(x)\ket{1}\bra{2}+h.c.\right)dx.
\end{align}
Here $d=L$ (R) represents left (right) propagating photons in the waveguide and $\hat{c}_{d}$ represents annihilation operators for the $d$th direction. The real-valued parameter $V_d$ quantifies the atom-photon interaction strength with the Dirac delta function specifying the location of the QE. Next, in the single excitation sector of the Hilbert space, we write the quantum state of the system as
\begin{align}\label{QS}
\ket{\Psi} = \Big[\sum_{d=L,R}\int\varphi_{d}(x)\hat{c}^{\dagger}_{d}dx+ \sum_{j=2,3} e_{j}\ket{j}\bra{1}\Big]\ket{\varnothing},
\end{align}
where $\ket{\varnothing}$ represents the ground state of the system (i.e. QE being unexcited and no photons in the waveguide). By inserting Eq.~(\ref{HamLin}) and Eq.~(\ref{QS}) into the time-independent Schr\"odinger equation $\hat{\mathcal{H}}\ket{\Psi}=\hbar\omega\ket{\Psi}$ we arrive at the following coupled equations for the amplitudes
\begin{subequations}
\begin{align}
    -iv_{g}\varphi_{R}(x) + V_{R}e_{2}\delta(x) &= \omega\varphi_{R}(x),\label{Linear_Eq_t} \\
    iv_{g}\varphi_{L}(x) + V_{L}e_{2}\delta(x) &= \omega\varphi_{L}(x),\label{Linear_Eq_r} \\
    V_{R}\varphi_{R}(0) + V_{L}\varphi_{L}(0) + \frac{\Omega}{2}e_{3} &= (\omega - \widetilde{\omega}_{2})e_{2} \label{Linear_Eq_e2}, \\
    \frac{\Omega}{2}e_{2} &= (\omega - \widetilde{\omega}_{3})e_{3}, \label{Linear_Eq_e3}
\end{align}    
\end{subequations}
where $\hbar\omega$ is the energy of the photon incident from the left end of the waveguide. Our aim now is to calculate the single-photon transmission and reflection probabilities and for that, we assume the left and right field amplitudes obey the following ansatzes
\begin{subequations}
\begin{align}
\varphi_{R}(x) &= e^{ikx}\Theta(-x) + te^{ikx}\Theta(x),\label{Phi_R_Sol} \\
\varphi_{L}(x) &= re^{-ikx}\Theta(-x).\label{Phi_L_Sol}
\end{align}    
\end{subequations}
Here $t$, $r$ are the transmission and reflection coefficients which are respectively related to the transmission and reflection probabilities through $|t|^2=T$ and $|r|^2=R$. With this the amplitude equations for $\varphi_{R/L}(x)$ becomes
\begin{subequations}
\begin{align}
iv_{g}(1-t) + V_{R}e_{2} &= 0, \\
-iv_{g}r + V_{L}e_{2} &= 0, \\
\frac{V_{R}}{2}(t+1)+\frac{V_{L}}{2}r+\frac{\Omega}{2}e_{3} &= (\omega - \widetilde{\omega}_{2})e_{2}.
\end{align}   
\end{subequations}
The solution of this set of coupled equations yields the transmission and reflection amplitudes for the linear ($l$) waveguide dispersion case as given by
\begin{align}
    t^{(l)} &= \frac{\left(\omega-\widetilde{\omega}_3\right)[\left(\omega-\widetilde{\omega}_2 \right) + i(\Gamma_{L} - \Gamma_{R})/4] - \Omega^2/4}{\left(\omega-\widetilde{\omega}_3\right)[\left(\omega-\widetilde{\omega}_2 \right) + i(\Gamma_{L}+\Gamma_{R})/4] - \Omega^2/4} \label{Linear_T_Amp}, \\
    \hspace{-2mm}r^{(l)} &= \frac{-i\sqrt{\Gamma_R\Gamma_L}/2 \left(\omega-\widetilde{\omega}_3\right)}{\left(\omega-\widetilde{\omega}_3\right)[\left(\omega-\widetilde{\omega}_2\right) + i(\Gamma_{L}+\Gamma_{R})/4] - \Omega^2/4} \label{Linear_R_Amp},
\end{align}
with $\Gamma_{d} \equiv \frac{2V_{d}^{2}}{v_{g}}$ being the emitter-waveguide coupling rate for the $d$th direction. We point out that these results have already been reported for the single three-level QE problem by Witthaut et al. and Mirza et al. \cite{witthaut2010photon, martens2013photon, mirza2018influence}. 
\begin{figure*}
\centering
\begin{tabular}{@{}cccc@{}}
\includegraphics[width=2.3in, height=1.6in]{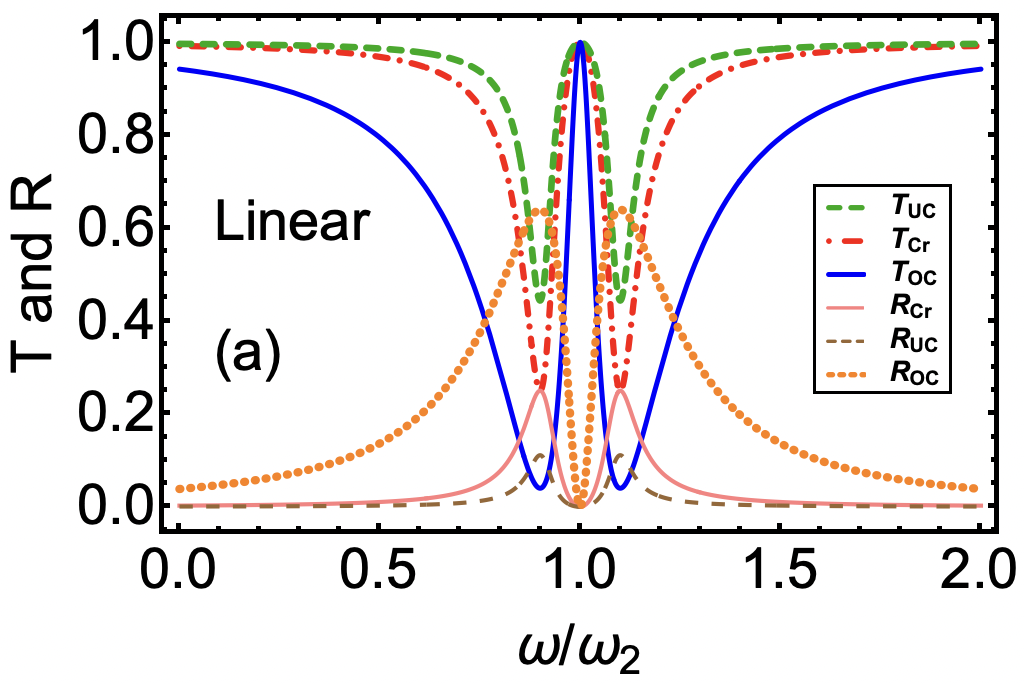} &
\includegraphics[width=2.3in, height=1.6in]{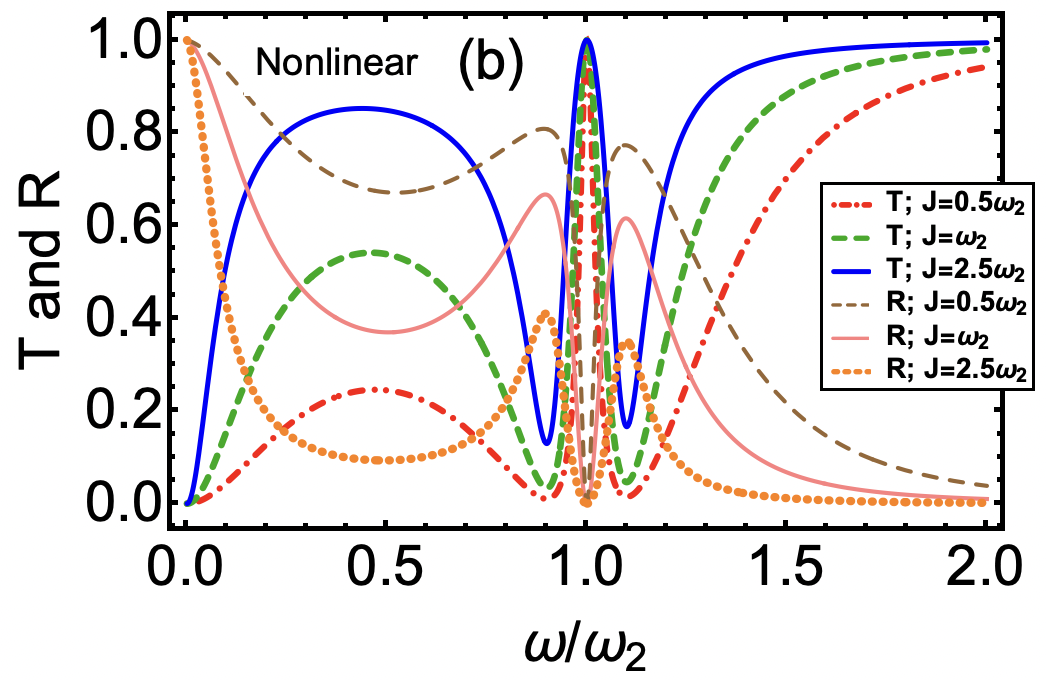} &
\includegraphics[width=2.3in, height=1.6in]{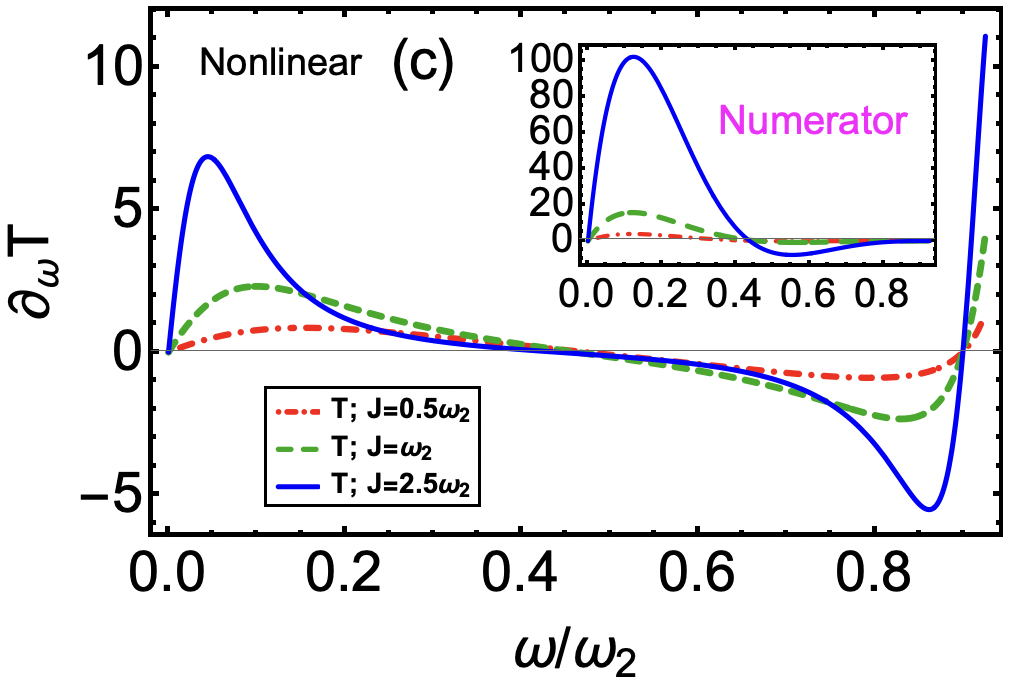} 
\end{tabular}
\vspace{-5mm}
\captionsetup{
format=plain,
margin=1em,
justification=raggedright,
singlelinecheck=false
}
\caption{(Color online) Single-photon T and R probability as a function of frequency $\omega$ for the single QE ($N=1$) problem in the (a) linear dispersion case with critical (${\rm Cr}$) coupling regime $\Gamma_d=\gamma_2$, under coupled (${\rm UC}$) regime $\Gamma_d=\gamma_2/2$, and over coupled (${\rm OC}$) regime $\Gamma_d=4\gamma_2$. (b) Nonlinear case. In all curves, we have selected an OC regime in which $\Gamma_d=4\gamma_2$ and varied values of $J$. (c) The derivative of $T$ with respect to photon frequency $\omega$ to examine the dependence of transmission plateau in the nonlinear case on different values of $J$. The inset plot curves show the numerator part of $\partial_\omega T$ again for different $J$ values. The common parameters used in all plots are $\Omega=0.2\omega_2, \gamma_2=0.1\omega_2$, $\gamma_3=0$, and $\Delta=0$.}\label{Fig2}
\end{figure*}
In Fig.~\ref{Fig2}(a) we plot the $T(\omega)$ and $R(\omega)$ for the linear dispersion case under the symmetric emitter-waveguide coupling assumption ($\Gamma_L=\Gamma_R=\Gamma$). By selecting experimentally feasible parameters inspired from Ref.~\cite{witthaut2010photon} we consider three cases based on how emitter-waveguide coupling rate $\Gamma$ compares with the decay rate $\gamma_2$. We find that in the ${\rm OC}$ case (solid blue curve) $T$ exhibits closest to the perfect EIT pattern where the maximum transmission occurs at $\omega=\omega_2$ and the linewidth of the transparency window is controlled by the Rabi frequency $\Omega$. Additionally, the shallowness and width of the two side dips (originating from the reflection and therefore, interference of the photon) occurring at $\omega=\pm \Omega/2$ under $\Delta=0$ and $\gamma_3=0$ conditions can be manipulated through the emitter-waveguide coupling parameter $\Gamma$. Finally, we indicate that in the OC regime, four points exist where $T$ and $R$ curves intersect allowing the preparation of single-photon quantum superposition states of transmission and reflection (see for instance, Ref.~(\cite{shen2005coherent, mirza2017chirality}) for a similar behavior for two-level QE wQED).

\begin{figure*}
\centering
\begin{tabular}{@{}cccc@{}}
\includegraphics[width=2.4in, height=1.7in]{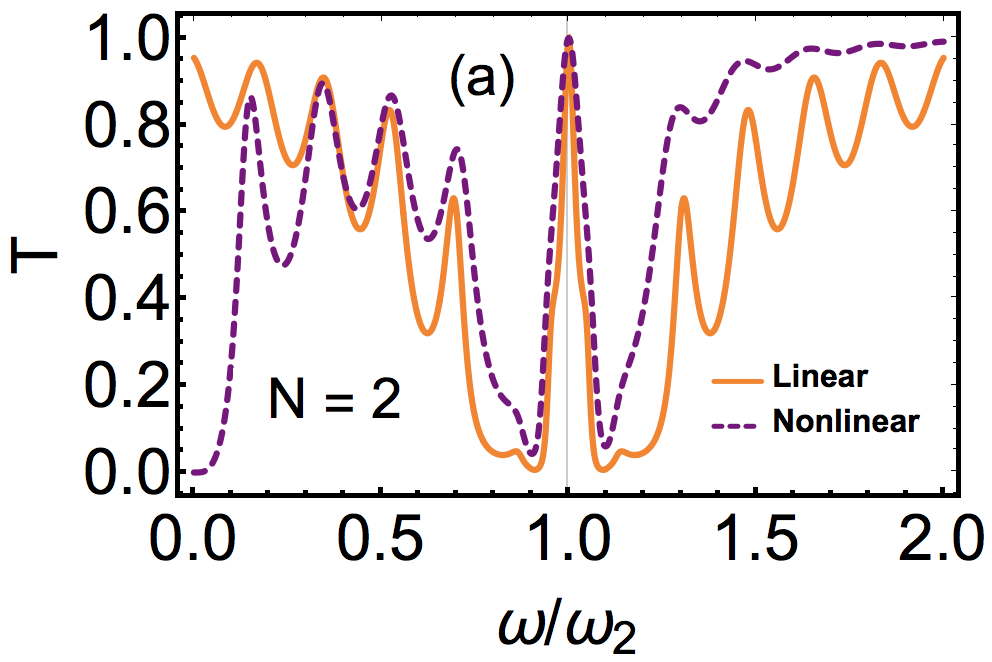} &
\includegraphics[width=2.25in, height=1.7in]{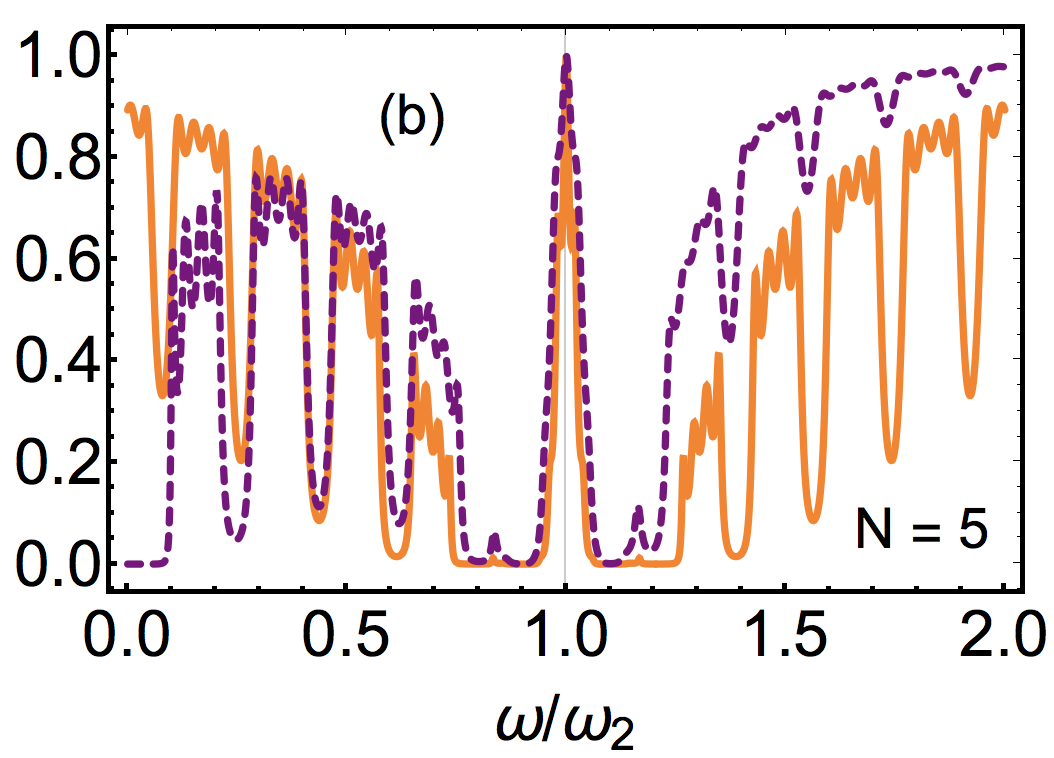} &
\includegraphics[width=2.25in, height=1.7in]{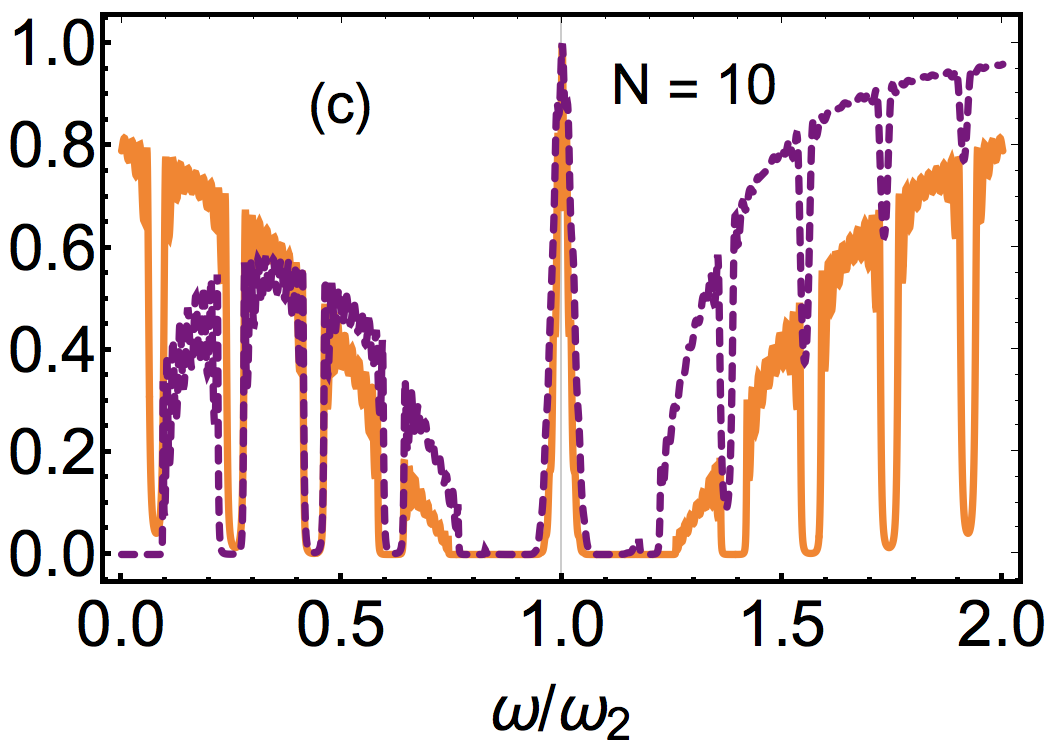} 
\end{tabular}
\vspace{-5mm}
\captionsetup{
format=plain,
margin=1em,
justification=raggedright,
singlelinecheck=false
}
\caption{(Color online) Single-photon transport properties for a periodic chain consisting of (a) two, (b) five, and (c) ten identical QEs coupled with the bidirectional waveguide. In each plot, both linear (solid orange curve) and nonlinear (dashed purple curve) cases have been plotted for comparison. In all plots, we have selected an OC regime with $\Gamma_d=4\gamma_2$ and considered a strong nonlinear dispersion with $J=2.5\omega_2$. The lattice constant $L$ has been selected to be $0.5\lambda_0$ where $\lambda_0=2\pi v_g/\omega_2$. The rest of the parameters are the same as used before.}\label{Fig7}
\end{figure*}
{\it Nonlinear Dispersion case: Single QE problem.}---We now turn our attention to the corresponding nonlinear dispersion regime for the single QE case. To this end, we consider a waveguide model consisting of a one-dimensional array of coupled Fabry-Per\'ot cavities (as shown in Fig.~\ref{Schematic_Diagram}(c))). Under the tight-binding approximation, such a model is known to generate a cosine dispersion relation of the form  $\omega(k) = \omega_J-2J\cos(kL)$ \cite{zhou2008controllable, talukdar2022two} (see Fig.~\ref{Schematic_Diagram}(d)), where $\omega_J$ represents the average $\omega_k$ value in the cosine curve. As derived in the Appendix, in comparison to the linear case, we describe the free waveguide Hamiltonian in Eq.(\ref{Hamiltonian_Linear}) in the nonlinear ($nl$) case as
\vspace{-1mm}
\begin{align}\label{Free_Wave_H_Nonlinear}
   \hat{\mathcal{H}}_{W}^{(nl)} = -2J\sum_{d=L,R}\int &\hat{c}^{\dagger}_{d}(x)
   \bigg(\sum_{j=0}^{\infty}\frac{\partial_{x}^{2j}}{(2j)!}\bigg)\hat{c}_{d}(x)dx,
\end{align}
where we set $\omega_J=0$. Proceeding further, we truncate the cosine dispersion to the quadratic level (as shown in Fig.~\ref{Schematic_Diagram}(d)). Following the same line of calculations as before we arrive at a set of four coupled equations. Only Eq.~(\ref{Linear_Eq_t}) and Eq.~(\ref{Linear_Eq_r}) change and take the new form as
\begin{align}
-J\partial_{x}^{2}\varphi_{d}(x) + V_{d}e_{2}\delta(x) &= \omega\varphi_{d}(x), \label{Nonlinear_Eq_t} 
\end{align}    
We note that the pre-factors of $iv_{g}$ appearing in the linear case with opposite signs for left and right directions (see Eq.(\ref{Linear_Eq_t}) and Eq.(\ref{Linear_Eq_r})) have now been swapped with second-order derivative term times the hopping rate with the same negative sign for both $d=L$ and $d=R$. Next, we express Eq.~(\ref{Nonlinear_Eq_t}) in terms of $t$ and $r$ as
\begin{align}
-\frac{iJ\omega}{v_{g}}(t-1)+V_{R}e_{2} = 0, \text{and}
-\frac{iJ\omega}{v_{g}}r + V_{L}e_{2} = 0.
\end{align}
Solving for the transmission and reflection coefficients in the nonlinear (nl) case yields
\begin{align}
&t^{(nl)} = \frac{(\omega-\widetilde{\omega}_3)[\omega (\omega-\widetilde{\omega}_2) + i(\Gamma_{L} - \Gamma_{R})/(4J)] - \omega\Omega^2/4}{(\omega-\widetilde{\omega}_3)[\omega (\omega-\widetilde{\omega}_2) + i(\Gamma_{L}+\Gamma_{R})/(4J)] - \omega\Omega^2/4}, \nonumber\\
&r^{(nl)} = \frac{-2i\sqrt{\Gamma_R\Gamma_L} (\omega-\widetilde{\omega}_3)/(4J)}{(\omega-\widetilde{\omega}_3)[\omega (\omega-\widetilde{\omega}_2) + i(\Gamma_{L}+\Gamma_{R})/(4J)] - \omega\Omega^2/4}. \label{nonlinear_r_exp}
\end{align}
shape remains intact, indicating the robustness of EIT in the OC regime. Secondly, as we increase the nonlinearity, a side peak began to emerge around $\omega\sim\omega_2/2$ point. Interestingly this peak tends to form a plateau as we approach the nonlinearity value of $J=2.5\omega_2$ value. As shown in Fig.~\ref{Fig2}(c), the behavior of this plateau can be analyzed by considering the derivative $\partial _\omega T$ (a mathematical expression not reported here due to complexity). We notice that when $\omega/\omega_2 < 1$ the denominator of $\partial _\omega T$ is strictly positive, therefore the numerator controls the shape of the function in this region. This numerator (which turns out to be polynomial in $\omega$ to the eighth power) is proportional to sums of $J^{2}$ and $J^{3}$ terms under the conditions listed in Fig.~\ref{Fig2}. Hence this combination of a squared and cubed dependence, the height and width of the plateaus are found to be strongly correlated with $J$. We would like to highlight that this low-frequency spectral feature is peculiar to the nonlinear case (with no counterpart in the linear case) and can be used to manipulate the photon transport properties in novel ways without disturbing the advantages of the EIT spectrum.

{\it Extension to many-emitter case: photonic bands.}---We now extend our single QE setup to the scenario in which a periodic chain of identical three-level QEs is coupled with bidirectional waveguide fields. For the single photon problem it is well-known that $t$ and $r$ coefficients between consecutive atoms can be adequately linked in terms of transfer matrices \cite{mostafazadeh2020transfer}. Such a transfer matrix consists of two parts, namely, a part representing the response of a QE ($\mathbf{M}_{QE}$), and a part showing the free propagation of the photon in the waveguide between two consecutive atoms ($\mathbf{M}_F$). Under time reversal symmetry restrictions $\mathbf{M}_{QE}$ and $\mathbf{M}_F$ can be generically written as \cite{mostafazadeh2020transfer, zhou2008controllable},
\begin{equation}
    \mathbf{M}_{QE} = 
    \begin{bmatrix}
        1/t^\ast & -r^\ast/t^\ast \\
        -r/t & 1/t
    \end{bmatrix}, ~\text{and}~
    \mathbf{M}_F =
    \begin{bmatrix}
        e^{iqx} & 0 \\
        0 & e^{-iqx}
    \end{bmatrix},
\end{equation}
where $t$ and $r$ can be transmission and reflection coefficients for linear or nonlinear problems and $q$ represents the resonant wavenumber i.e. $q=\omega_2/v_g$. The free propagation introduces time delays in the problem which are necessary to distinguish between Markovian and non-Markovian regimes of wQED \cite{carmele2020pronounced}. Proceeding further we concentrate on the Markovian regime and divide the many QE problem into $N$ segments/blocks where each block consists of a single QE and a free propagating region leading to the form of the transfer matrix of a single block as $\mathbf{M}_B = \mathbf{M}_{QE}\times\mathbf{M}_F$.

For a chain of identical QEs, by applying Chebyshev's identity \cite{yeh1977electromagnetic} one can arrive at an expression for the net transmission $T_N$ under the no-loss scenario as
\begin{align}
    T_N=\left(1+\frac{|r^2|}{|t^2|}\frac{\sin^2 (NqL)}{\sin^2 (qL)}\right)^{-1}.
\end{align}
In Fig.~\ref{Fig7} we plot the net $T_N(\omega)$ for a chain of up to $N=10$ QEs coupled with a waveguide with linear and nonlinear dispersion. Selecting lattice constant $L=0.5\lambda_0$, we find the formation of photonic bands in both linear and strongly nonlinear cases ($J=2.5\omega_2$) where the difference between the two cases is most pronounced around $\omega\sim 0.25\omega_2$ region. However, in both cases, the main features of the EIT peak remain more or less undisturbed. It is known that such bands are formed due to interference between transmitted and reflected amplitudes from each QE boundary which for a large QE number reduce to a sharper frequency combs pattern with applications in atomic clocks, spectroscopy, and metrology \cite{liao2016single}.

\begin{figure}
\centering
\begin{tabular}{@{}cccc@{}}
\includegraphics[width=1.6in, height=2.5in]{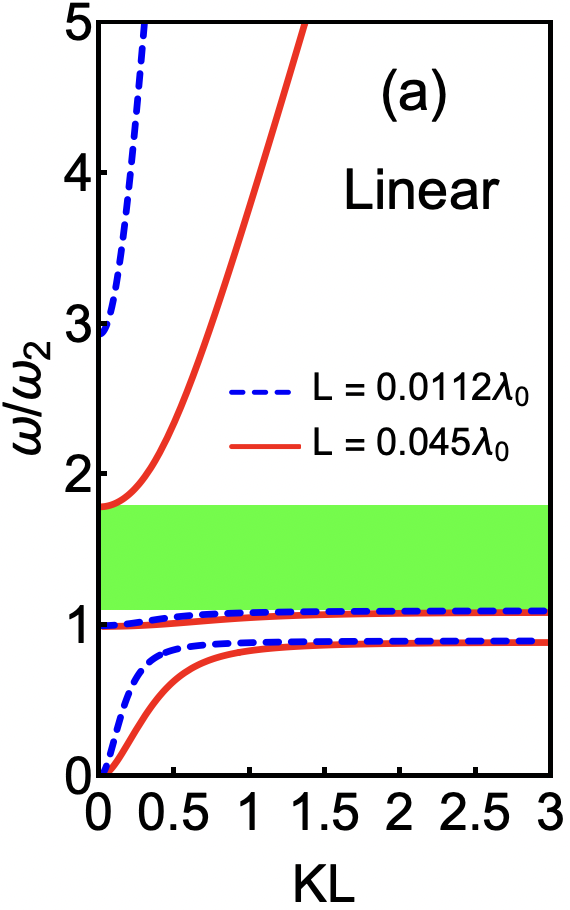} &
\includegraphics[width=1.5in, height=2.5in]{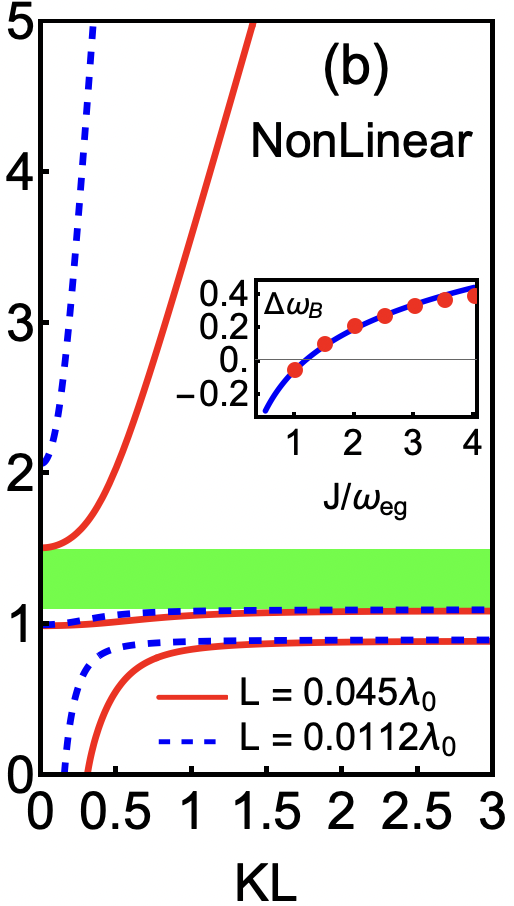} 
\end{tabular}
\vspace{-4mm}
\captionsetup{
format=plain,
margin=1em,
justification=raggedright,
singlelinecheck=false
}
\caption{(Color online) Dispersion curves for a (a) linear, and (b) nonlinear waveguide QED setup. In each plot, two closely spaced inter-atomic separations are selected. The inset in (b) indicates the difference between bandgap $\Delta\omega_B$ for linear and nonlinear cases plotted against the nonlinearity parameter $J$ (red dots are points indicating the difference in the width of the bandgaps at specific $J$ values while the blue curve is a log fit as discussed in the text). The rest of the parameters are as used in Fig.~\ref{Fig7}.}\label{Fig8}
\end{figure}

To examine the dispersion characteristics of the photon we consider an infinitely long QE chain with lattice constant $L$. Applying the Bloch theorem \cite{shen2007stopping} with $K$ being the Bloch vector we obtain
\vspace{-2mm}
\begin{align}
    \cos(KL)=\frac{1}{2}{\rm tr}\lbrace {\bf M}_{B} \rbrace = \mathfrak{Re}\left[\frac{e^{-iqL}}{t}\right],
\end{align}
Inserting the transmission coefficients $t^{(l)}$ and $t^{(nl)}$ in the last equation we arrive at the following dispersion relations for linear and nonlinear case as
\begin{subequations}
    \begin{align}
    &(l):\cos(KL) = \cos(qL) + \sin(qL)\left\lbrace \frac{\Gamma\delta_3}{2\Lambda^2} \right\rbrace ,\\
    &(nl):\cos(KL) = \cos(qL)+ \sin(qL)\left\lbrace \frac{\Gamma\delta_3}{\overline{\Lambda}^3 J} \right\rbrace,
\end{align}
\end{subequations}
where $\Lambda^2=\delta_2\delta_3-\Omega^2/4$ and $\overline{\Lambda}^3=\omega(\delta_2\delta_3-\Omega^2/4)$ with $\delta_j=\omega - \omega_j$, $\forall j=2,3$. In Fig.~\ref{Fig8} we plot the dispersion curves for linear and nonlinear cases. For simplicity, we have selected $\gamma_2=\gamma_3=0$ and $J=2.5\omega_2$. Furthermore, we have chosen $L$ to be much smaller than the characteristic wavelength $\lambda_0$ to make forbidden bands more visible. As the main result, we observe that nonlinearity can be used to control the width of the forbidden bands. For instance, a nonlinearity of $J=2.5\omega_2$ can reduce the width of the forbidden band (green shaded region) by a factor of $1/2$ as compared to the corresponding linear case. 

The difference in the width of the photonic band gap introduced by the nonlinearity can be quantified through the function $\Delta\omega_B \equiv \omega_{lB} - \omega_{nlB}$, where $\omega_{lB/nlB}$ is the value at which the band gap ends after $\omega/\omega_{2} = 1$. Due to the transcendental nature of the equations such a band gap can be approximated by the following relationship, $\Delta\omega_B(J) \approx \omega_2\log_{b}(J/\omega_2) + \xi$ (see inset of Fig.~\ref{Fig8}(b)). For the case of $L = 0.045\lambda_{0}$, b$\sim 16.751$ and $\xi\sim -0.047\omega_2$ suffice. This approximate behavior shows that the difference in the band gap grows as a function of $J$. However, as the difference obeys a logarithmic dependence, there also exists a range of J values for which the nonlinear band gap can also take a larger value range than the linear band gap (for instance, for $L = 0.045\lambda_{0}$ we found all $J\leq\sim 1.141$ leads to  $\Delta\omega_B \leq 0$).

{\it Summary and Conclusions.}---In this work, we have examined the single photon transport properties in three-level QEs chains coupled with 1D waveguides with linear and nonlinear dispersions. In the single-emitter case, we found that linearly dispersive waveguides lead to the standard EIT spectrum in the OC regime. In the nonlinear case, at the quadratic level, we found that the EIT profile prevails as we increase the strength of the nonlinearity parameter of dispersion. However, for smaller frequencies, we observed the formation of a considerably higher plateau in the transmission spectrum. 

In the many-emitter OC case, we noticed the formation of band structures where EIT spectral features remain intact. Again the difference between the linear and nonlinear dispersions was most pronounced for smaller frequencies. Finally, the photonic band gaps were investigated using the Bloch theorem where we mainly found the narrowing of band gaps for elevated values of nonlinearity for smaller inter-emitter separations. Our results revealed that nonlinear wQED provides new ways of controlling the transport properties of single photons with possible applications in quantum memory-enabled long-distance quantum communication protocols. 

{\it Acknowledgements.}---This work is supported by the NSF Grant \# LEAPS-MPS 2212860 and the Miami University College of Arts and Science \& Physics Department start-up funding.
\vspace{-5mm}

\setcounter{equation}{0}
\renewcommand\theequation{A.\arabic{equation}}
\section*{APPENDIX: Derivation of $\mathcal{H}^{(nl)}_W$}
The derivation of the free waveguide Hamiltonian for the nonlinear case begins with the multimode quantum harmonic oscillator model of the waveguide which for the right direction takes the form
\begin{align}
\sum_{k}\omega_{k_{R}}\hat{a}_{k_{R}}^{\dagger}\hat{a}_{k_{R}} = & -\frac{J}{\pi}\iiint\hat{c}_{R}^{\dagger}(x)\hat{c}_{R}(x')\notag\\
&\times\cos(k_{R})e^{ik_{R}(x-x')}dxdx'dk_{R}.\label{Appendix_Ref_1}
\end{align}
Utilizing the Taylor series expansion of the cosine function i.e. $\cos(k_{R}) = \sum_{j=0}^{\infty}\frac{(-1)^{j}k_{R}^{2j}}{(2j)!}$, the term under the integral reads as 
\begin{align}
\cos(k_{R})e^{ik_{R}(x-x')} 
=  \Bigg[\sum_{j=0}^{\infty}\frac{\partial_{x}^{2j}}{(2j)!}\Bigg]e^{ik_{R}(x-x')}.\label{Appendix_Ref_2}
\end{align}
Here we have set lattice constant $L=1$. Taking the expansion of cosine as defined in the last equation and inserting it into the right-hand side of Eq.(\ref{Appendix_Ref_1}) results in the following integral 
\begin{align}
-\frac{J}{\pi}\iiint&\hat{c}_{R}^{\dagger}(x)\hat{c}_{R}(x')\Bigg[\sum_{j=0}^{\infty}\frac{\partial_{x}^{2j}}{(2j)!}\Bigg]e^{ik_{R}(x-x')}dxdx'dk_{R},
\end{align}
where integrating over $k_{R}$ and $x'$ and going through a similar calculation for the left direction will then result in the $\mathcal{H}^{(nl)}_W$ as expressed in Eq.(\ref{Free_Wave_H_Nonlinear}).


\bibliographystyle{ieeetr}
\bibliography{paper}
\end{document}